\documentclass[
 reprint,
 amsmath,amssymb,
 aps,
]{revtex4-2}

\usepackage{graphicx}
\usepackage{dcolumn}
\usepackage{bm}
\usepackage{dsfont}
\usepackage{amsthm,amsmath,amssymb}
\usepackage{exscale}
\usepackage{relsize}
\usepackage{xcolor}
\usepackage{physics}
\usepackage{mathrsfs}

\begin{document}

\preprint{APS/123-QED}

\title{Intense, wideband optical waveform generation by self-balanced amplification of fiber electro-optical sideband modulation}

\author{Yuzhuo Wang$^1$}
\email{zhuodashi@163.com}
\author{Yizun He$^1$}
\author{Lingjing Ji$^1$}
\author{Jiangyong Hu$^1$}
\author{Xing Huang$^1$}
\author{Yudi Ma$^1$}
\author{Liyang Qiu$^1$}
\author{Kaifeng Zhao$^{2,3}$}
\author{Saijun Wu$^1$}
\email{saijunwu@fudan.edu.cn}

\affiliation{$^1$Department of Physics, State Key Laboratory of Surface Physics and Key Laboratory of Micro and Nano Photonic Structures (Ministry of Education),Fudan University, Shanghai 200433, China.\\
$^2$Key Laboratory of Nuclear Physics and Ion-Beam Application (MOE), Fudan University, Shanghai 200433, China.\\
$^3$Institute of Modern Physics, Department of Nuclear Science and Technology, Fudan University, China.
}

\begin{abstract}
  We demonstrate a simple method to obtain accurate optical waveforms with a GHz-level programmable modulation bandwidth and Watt-level output power for wideband optical control of free atoms and molecules. Arbitrary amplitude and phase modulations are transferred from microwave to light with a low-power fiber electro-optical modulator. The sub-milliWatt optical sideband is co-amplified with the optical carrier in a power-balanced fashion through a tapered semiconductor amplifier (TSA). By automatically keeping TSA near saturation in a quasi-continuous manner,  typical noise channels associated with pulsed high-gain amplifications are efficiently suppressed. As an example application, we demonstrate interleaved cooling and trapping of two rubidium isotopes with coherent nanosecond pulses.
\end{abstract}
 
 \maketitle
 
 \section{Introduction}

Optical control of atomic motion is traditionally accomplished by weakly dressing atoms in their ground-state  manifolds, such as laser cooling, atom interferometry, and ion-based quantum information processing~\cite{MetcalfBook,Blatt2008, Cronin2009,Entanglement2013,Barry2014,Moses2017, Kozyryev2017}. The long coherence time associated with the weakly dressed ground states makes it possible to precisely control the dynamics using modulated CW lasers, typically through acousto-optical modulation (AOM) with a MHz-level bandwidth~\cite{Thom2013}.  On the other hand, full control of the strong-transition dynamics between ground and excited states becomes an emergent scenario, with many applications in atomic and quantum optics, such as for ultrafast optical acceleration of spinnor mattterwave~\cite{Miao2007,Jayich2014,Heinrich2018, Long2019, He2020b}, precise control of light-assisted interactions~\cite{Koch2012, Carini2015}, and to access subradiant physics~\cite{Scully2015,Facchinetti2016, He2020a}.  
Since strong transitions have coherence time radiatively limited to tens of nanoseconds, their full and coherent control requires optical waveforms with modulation bandwidth at the GHz level beyond standard CW modulation technology. Although ultrafast pulses can have bandwidth beyond THz, the pulse spectral brightness is usually too weak to efficiently drive the narrow transitions~\cite{Goswami2003, Zhdanovich2008, Ma2020}.

   
Efforts have been made to generate intense, coherent optical waveforms with GHz modulation bandwidth for atomic physics applications~\cite{Gould2015}. 
For example, 
coherent pulse trains with short inter-pulse delays are generated in the time domain to excite atoms efficiently ~\cite{Heinrich2018,Ma2020}. Fiber electro-optical modulators (fEOM) with $\sim$10~GHz bandwidths are exploited to transfer modulation from microwaves to light~\cite{Gould2015,Kaufman2017,Xuejian2017,He2020a,Clarke2021,Macrae:21}. Compared with shaping ultrafast pulses, modulation of CW lasers is more convenient for achieving long coherence time for, {\it e.g.}, complex optical control with composite pulses~\cite{He2020a,He2020b,He2021a}.
However, the integrated lithium-niobate-based fEOM suffers from severe photo-refractive damage at short wavelengths, limiting the throughput to a few tens of milliwatts or less in atomic physics applications~\cite{Gould2015,Kaufman2017,Xuejian2017,He2020a,Clarke2021}. The weak signal could be amplified into a watt-level output using tapered semiconductor amplifiers (TSA) under a double-pass configuration~\cite{Bolpasi2010,Gould2015,Orrest2019}.  However, unlike continuous seeding~\cite{Bolpasi2010,Clarke2021}, when the seeding waveform is modulated in amplitude, the amplified spontaneous emission (ASE) can be severe. The ASE problem is partially addressed in ref.~\cite{Gould2015}  by carefully managing the optical gain in both time and frequency domain, although the remaining ASE that shares the time-frequency window with the amplitude-modulated waveform output can still be detrimental. In refs.~\cite{Xuejian2017,He2020a} ASE is reduced by amplifying a CW laser first and then pulsing the high power output into the fEOM with a low enough duty cycle for the wideband modulation without damaging the fEOM. Although pulsed waveforms with hundreds of milliwatts of power can be achieved this way, the procedure precludes the possibility of generating continuous waveforms with high average power.  Apart from the ASE problem, it is known that when amplifying amplitude-modulated light, the transient change of the gain saturation level in TSA associated with the electron-hole density leads to self-phase modulation (SPM) ~\cite{AgrawalGovindP1989,Hong1994, Cruz2006, Baveja2010,Luo2013,Meng2018} and distortion of individual output waveforms.
  
In this work, we introduce a simple method to achieve intense, wideband programmable optical waveforms with substantially suppressed noise associated with ASE and SPM effects. The method starts with phase-modulating a CW laser with fEOM at a microwave carrier frequency $f_{\rm c}$. Complex waveforms are transferred from the microwave to the optical sidebands of the fEOM output, which is then filtered and amplified with TSA. To suppress the SPM waveform distortion and ASE noise, we tune the optical filter to balance the power between the desired sideband (when the fEOM microwave modulation is on) and the optical carrier (when the modulation is off) so as to maintain a nearly constant seeding power and a consistent level of TSA saturation. The optical carrier off-resonant to the atomic transition can be subsequently filtered away. We demonstrate the method with a wideband optical waveform generation system with $\sim 1$~W output power and $\omega_{\rm M}\sim 2\pi\times 4$~GHz programmable waveform modulation bandwidth for fast cooling and control of rubidium isotopes.

In the following, we first outline the operation principle of the amplified optical waveform generation system. We then detail the performance of our rubidium laser system and present an example application of the system for cooling and trapping with interleaved nanosecond pulses.

  \section{Methods }
  
  \subsection{Sideband modulation}

  \begin{figure}
    \centering
    \includegraphics[width=0.5\textwidth]{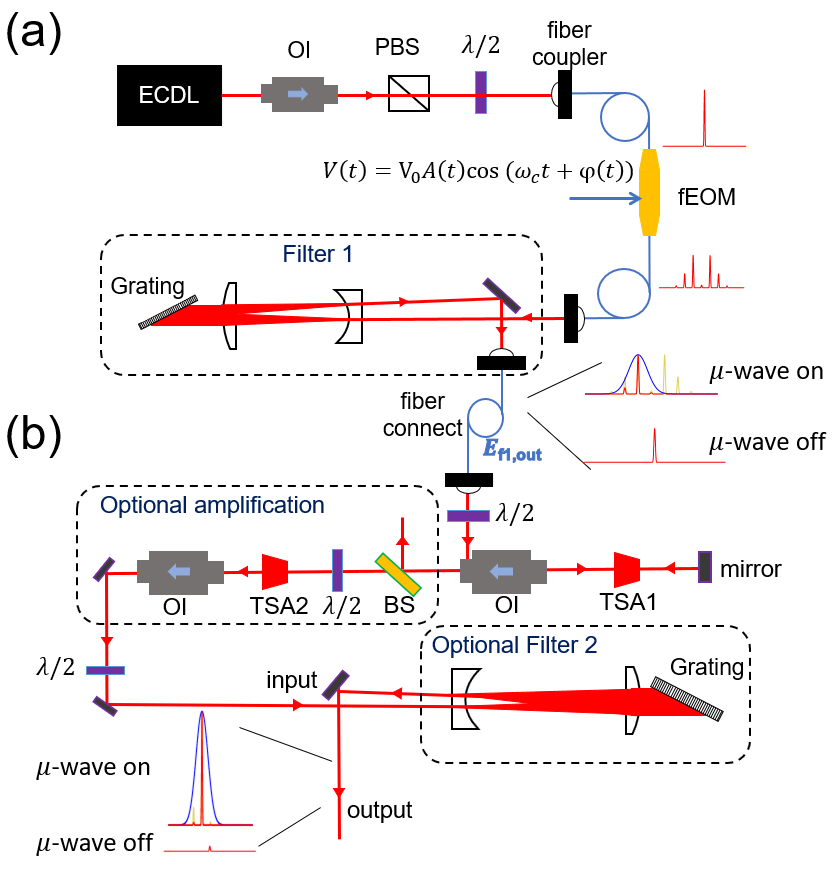}
        \caption{ The schematic setup of the waveform generation system. The spectrum of the optical waveform is  illustrated at each stage of the amplified modulation. (a) Schematic diagram of fEOM modulation and first optical filtering. CW laser from ECDL is modulated by fEOM with a programmable microwave signal. The fEOM output is collimated into a suitable size and filtered by grating diffraction before being coupled into a single-mode fiber. (b) High-gain optical amplification. The TSA$_1$ is seeded from the side-port of an optical isolator for double-pass amplification. Part of the output power seeds TSA$_2$ for an optional additional amplification. An optional Filter$_2$ serves to remove the optical carrier from the final output. ECDL: External cavity diode laser. O.I.: Optical isolator. PBS: Polarization beam splitter. BS: Beam splitter.}\label{fig:setup-filer}
  \end{figure}

  The setup of our laser system is schematically illustrated in Fig.~\ref{fig:setup-filer}.  A frequency-stabilized CW laser (an external-cavity-diode laser, ECDL) is phase-modulated through fEOM by an amplitude and phase-modulated microwave signal $V(t)=V_0 A(t)\cos( \omega_{\rm c} t+\varphi(t))$. Here $\omega_{\rm c}=2\pi f_{\rm c}$ at the 10~GHz level is the microwave carrier frequency. $A(t)\in [0, 1]$ and phase $\varphi(t)\in [0, 2\pi)$ are the amplitude and phase modulation functions to be transferred to light, respectively. We use complex $E_{\rm in}$ to represent the optical field. The output from fEOM can be expressed as:
  
  \begin{equation}
    \begin{array}{l}
    E_{\rm fEOM,out}(t)=e^{i \theta_0 A(t) \cos ( \omega_{\rm c} t +\varphi(t))}E_{\rm in}(t) \\
    ~~~~~~=\sum_n i^n J_n(\theta_0 A(t)) e^{i n ( \omega_{\rm c} t+\varphi(t))} E_{\rm in}(t).
    \end{array}\label{eqEOM}
    \end{equation}
  As described by the 2nd line of Eq.~(\ref{eqEOM}), the phase-modulated output can be decomposed into an array of optical sidebands, with $n^{\rm th}$-order $E_{n,{\rm out}}(t)=C_n(t) E_{\rm in}(t)$ subjected to complex modulation function $C_n(t)=J_n(\theta_0 A(t)) e^{i n\varphi(t)}$. Here $J_n$ is the $n^{\rm th}$-order Bessel function. The maximum electro-optical phase shift $\theta_0=\beta V_0$ is determined by the peak microwave voltage $V_0$ and the fEOM phase response coefficient $\beta$. By adjusting $V_0$ to let $\theta_0=\theta_{n_0}\sim n_0+1$, the magnitude of $|J_{n_0}(\theta_0)|^2$ can be maximized to the $n_0^{\rm th}$-order maximum as $J_{n_0,{\rm max}}^2$ for efficient generation of an $n_0^{\rm th}$ order sideband. To transfer a specific complex modulation to the $n_0^{\rm th}$ order sideband, {\it i.e.}, for $C_{n_0}(t)=a(t)e^{i\phi(t)}$, one simply programs the rf waveform as 
  \begin{equation}
    \begin{array}{l}
  A(t)=J_{n_0}^{-1}(a(t) J_{n_0,{\rm max}})/\theta_{n_0}\\
  \varphi(t)=\phi(t)/n_0.
    \end{array}\label{eq:invB}
  \end{equation}


  \subsection{The first optical filter}
  We send the fEOM output $E_{\rm fEOM,out}$ through an optical filter (Filter$_1$ in Fig.~\ref{fig:setup-filer}) to select the desired $n_0^{\rm th}$-order sideband and to attenuate the  optical carrier ($n=0$) to a suitable level. Here we consider the simplest example of optical filtering by grating diffraction. A grating constant $d<2 \lambda$ is preferred to achieve good diffraction efficiency near the Lithrow angle $\alpha \approx {\rm arcsin}(\lambda/2d)$~\cite{White2012}. With the filter input expanded into a Gaussian beam profile with Gaussian waist $w$, and the diffraction order mode-matched by the single-mode fiber, the fiber output power is approximately filtered in frequency by 
  \begin{equation}
  H(\omega)=\eta e^{-(\omega-\omega_{\rm A})^2/\Delta \omega_1^2}. \label{EqGaussFilter}
  \end{equation}
  Here $\eta$ is  the maximum filter efficiency. The central frequency $\omega_{\rm A}=\omega_{\rm L}+n_0 \omega_{\rm c}$ resonant to an atomic transition frequency is achieved by offset-locking the ECDL frequency $\omega_{\rm L}$ and tuning the grating angle $\alpha$ to maximize $H(\omega_{\rm A})$. By adjusting the laser beam polarization, a $\sim 70\%$ grating diffraction efficiency can be achieved in the near-infrared region, leading to a typical $\eta\approx 40\%$ overall efficiency after fiber coupling losses. The filter bandwidth $\Delta \omega_1=\frac{2\pi c}{w  {\rm tan}\alpha} = \frac{2\pi c}{w}\sqrt{4d^2/\lambda^2-1}$ can be adjusted with $w$ to match
  \begin{equation}
  H( \omega_{\rm L})/H(\omega_{\rm A})\approx J_{n_0,{\rm max}}^2.
  \label{eq:Hratio}
  \end{equation}
  As such, the resulting output $E_{\rm f1, out}$ is quasi-continuous  during the full-contrast $A(t)$ modulation with approximately constant average power. 
  
  \subsection{Self-balanced amplification\label{secAmp}}

  
At near-infrared, to avoid photo-refractive damage, the fEOM throughput is limited to less than tens of milliwatts. With a sideband modulation efficiency limited to $J_{n_0,{\rm max}}^2$ (Eq.~(\ref{eqEOM})) and after the Filter$_1$ loss $\eta$ (Fig.~\ref{fig:setup-filer}a, Eq.~(\ref{EqGaussFilter})), the filtered $E_{\rm f1, out}$ is typically less than a milliwatt.  The weak signal can be amplified into a watt-level output using tapered semiconductor amplifiers under a double-pass configuration~\cite{Bolpasi2010,Gould2015,Orrest2019}. The ASE associated with the high gain can be efficiently suppressed by saturating the TSA gain with a strong enough constant seeding~\cite{Bolpasi2010}. Here, for the amplification of seeding $E_{\rm f1, out}$ with a time-dependent amplitude, the optical gain $g(t)$ and the effective index $n(t)$ of TSA is expected to display time-dependent dynamics associated with semiconductor carrier density~\cite{AgrawalGovindP1989,Hong1994, Cruz2006, Baveja2010,Luo2013,Meng2018}, leading to time-dependent ASE and SPM to severely degrade the quality to the amplified $E_{\rm f1, out}$.

In this work, we realize that for a high enough microwave modulation carrier frequency $\omega_{\rm c} \gg 1/\tau_{\rm c}$,  where $\tau_{\rm c}$ is the effective relaxation time of carrier density~\cite{AgrawalGovindP1989,Hong1994,Baveja2010}, the TSA carrier density can hardly follow the microwave modulation, but stay at a saturation level determined by the average seeding power. Therefore, when the quasi-continuous seeding maintains a nearly constant average power during the full $A(t)$ modulation in a self-balanced fashion, ASE can be suppressed in a way similar to the case of CW seeding~\cite{Bolpasi2010}, while SPM can be suppressed in a way similar to those achieved in multi-sideband seeding with large enough frequency intervals~\cite{Luo2013,Meng2018}. 
  


  \subsection{The second optical filter}
  
  With the microwave carrier frequency $\omega_{\rm c}$ at the 10~GHz level, the amplified quasi-continuous $E_{\rm out}$ from TSAs can already be directly applied to resonantly drive atomic systems with its $n_0^{\rm th}$ order sideband. However, the off-resonant optical carrier and additional sidebands could  be detrimental to certain applications, such as for resonant imaging applications. To further select the $n_0^{\rm th}$-order sideband from the quasi-continuous output, an optional Filter$_2$ with bandwidth $\Delta \omega_2\ll |\omega_{\rm c}|$ can be constructed with a grating filter (Fig.~\ref{fig:setup-filer}b)~\cite{White2012,He2020a}, or with other types of narrow-line filters~\cite{Palittapongarnpim2012}.
  

  \section{The rubidium laser system}
\begin{figure}[tbp]
  \centering
  \includegraphics[width=0.5\textwidth]{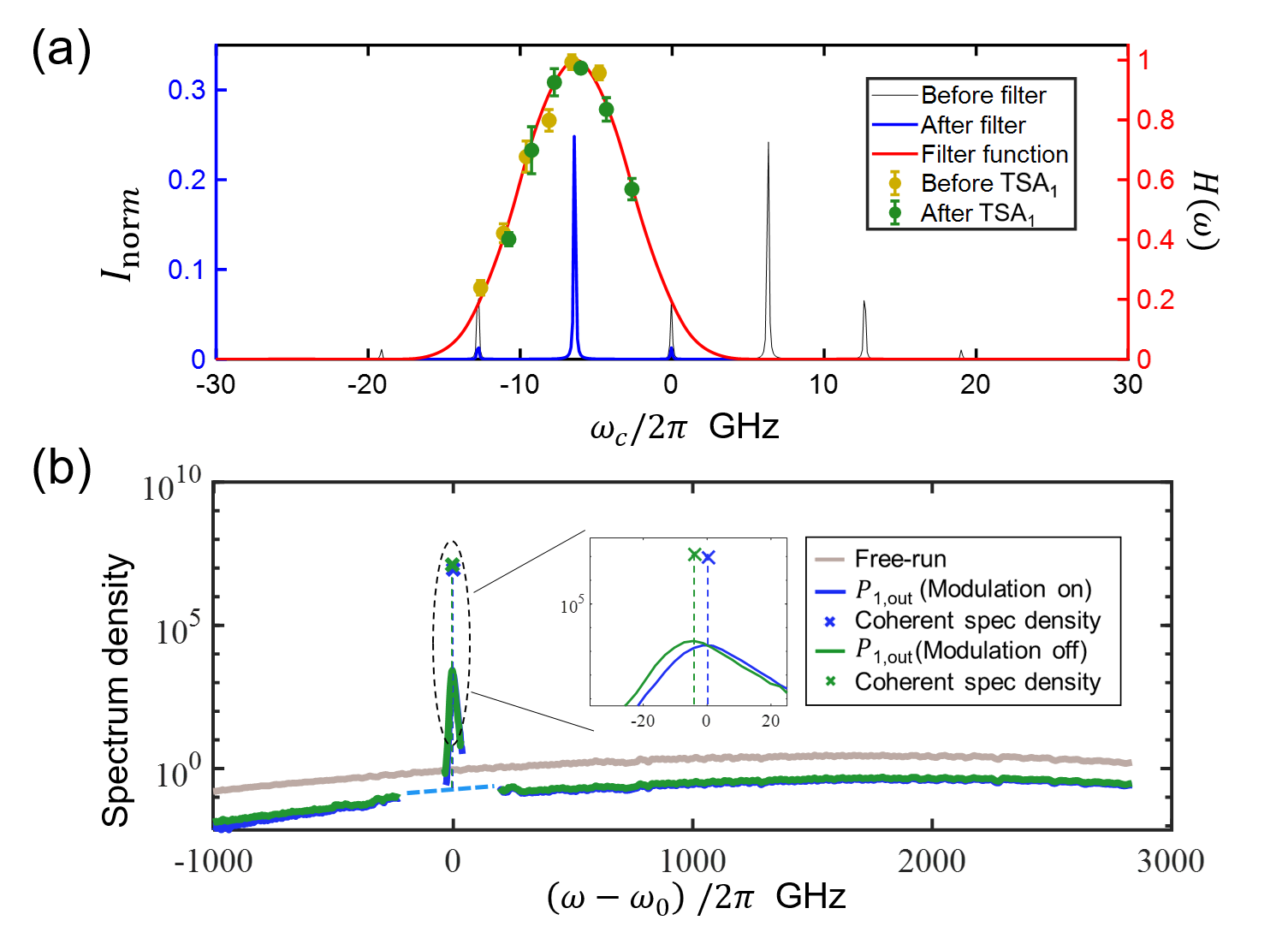}
  \caption{\label{fig:LF} (a) Bandwidth measurement for the $n_0=-1$ order output when the microwave carrier frequency $\omega_{\rm c}$ is sampled between $4-14$~GHz. The data points are for $n_0=-1$ order outputs before and after TSA$_1$ amplification, identified by a F-P spectrometer and normalized by the peak values. The EOM phase modulation depth $\theta_0\approx 1.8$ is optimized for the $n_0=-1$ order. The black and blue lines are $E_{\rm fEOM,out}$ and $E_{\rm f1,out}$ according to a numerical model of the grating-based optical filtering. The red curve is the Filter$_1$ function $H(\omega)$. (b) The spectrum of TSA$_1$ output measured with a grating spectrometer. The `X' markers  estimate the peak spectrum density according to the linewidth of the coherently amplified laser output.}
\end{figure}

So far we have outlined the general operation principle and the key elements of the amplified laser system. In the following, we provide additional details of the laser system designed for cooling and coherent manipulation of $^{85}$Rb,  $^{87}$Rb isotopes.  Here, to cover the needed frequency range spanned by the $^{87}$Rb ground-state hyperfine splitting, $f_{87,\rm hfs,g}=6.83$~GHz, we lock the ECDL to the $F=1-F'=0,1$ cross-over peak of the saturation spectroscopy. Part of the output can be shift directly with AOM to address the $F=1-F'=2$ repumping transition. The majority of the ECDL output is then modulated by fEOM with a carrier frequency of $\omega_{\rm c}=2\pi\times (f_{87,\rm hfs,g}+\Delta f)=2\pi\times 6.36$~GHz, with the $n_0=-1$ sideband to resonantly address the $F=2-F'=3$ transition of $^{87}$Rb. The $\Delta f=-0.47$~GHz is determined by the ECDL locking point as well as additional AOMs for the cooling laser control. We correspondingly set Filter$_1$ with a bandwidth $\Delta \omega_1\approx 6~$GHz. The filter function $H(\omega)$ is estimated in Fig.~\ref{fig:LF}a by measuring the $n_0=-1$ order output with a Fabry-Perot (F-P) spectrometer.


With 30~mW output from the ECDL, we obtain $P_{\rm f1}\approx 1~$mW of the filtered output $E_{\rm f1,out}$ to seed TSA$_1$. To fully utilize the $E_{\rm f1,out}$ for the double-pass seeding without optical damage, we limit the TSA$_1$ driving current to $I_1=1.0A$ to obtain $P_1=140$~mW  only. A 27~mW output is fiber-coupled to seed TSA$_2$ in the standard single-pass configuration. To allow the TSA$_2$ output to repump the $^{87}$Rb $F=1$ atoms, an AOM controlled $\sim$0.5~mW of ECDL output is also combined to seed TSA$_2$. By operating TSA$_2$ at $I_2=1.7A$, a $P_2\approx 720$~mW output is obtained, which is directly applied to the laser cooling and coherent control experiment without being filtered by the optional Filter$_2$. 

\subsection{ASE suppression}


We use a home-made wavelength-meter~\cite{White2012} to characterize the ASE background for both the TSA$_1$ and TSA$_2$ outputs. For this purpose, the laser power spectrum is measured under three operation modes: by $E_{\rm f1, out}$ seeding under full modulation, by $E_{\rm f1, out}$ seeding without modulation (with full power in 0$^{\rm th}$ order), and with TSA free-running (no seeding). We expose the camera twice in each measurement to separately retrieve the spectrum density of the coherent output and the ASE background. The background fluorescence is measured with a 100~ms exposure time, while the central coherent spectrum following the injected laser is measured with a 7~$\mu$s exposure time. The two spectrum data are then combined in Fig.~\ref{fig:TA2F}b after being relatively normalized by the ratio of the exposure time. Here the $\sim 3~$GHz frequency resolution of the spectrometer is beyond the  $~4$~MHz laser linewidth in this work, as verified by independent measurements. We normalize the spectrum density according to the laser linewidth to obtain the peak spectrum density relative to the ASE background. The spectrum density of the coherent output is $\sim 75$~dB above the ASE background at the seeding frequency which composes $\xi_1=90\%$ of the $P_1=140$~mW total output power.  Compared with TSA$_1$ in the free-running mode, the $E_{\rm f1,out}$ seeding leads to $\sim10$~dB suppression of ASE background, similar to Ref.~\cite{Bolpasi2010}. As discussed in Sec. IIB, with the Filter$_1$ bandwidth $\Delta \omega_1 \approx \omega_{\rm c}$, the $E_{\rm f1, out}$ seeding maintains the average seeding power $P_{\rm f1}\approx 0.6~$mW  during fEOM modulation, leading to a nearly identical level of ASE suppression regardless of the $E_{\rm f1, out}$ modulation strength. Similar ASE suppression is also obtained for TSA$_2$, with the coherent output  composing $\xi_2=84\%$ of the 720~mW total power and is $\sim 70$~dB beyond the ASE background in spectral density.

\subsection{Accurate waveform generation}

\begin{figure}[htbp]
  \centering
  \includegraphics[width=0.5  \textwidth]{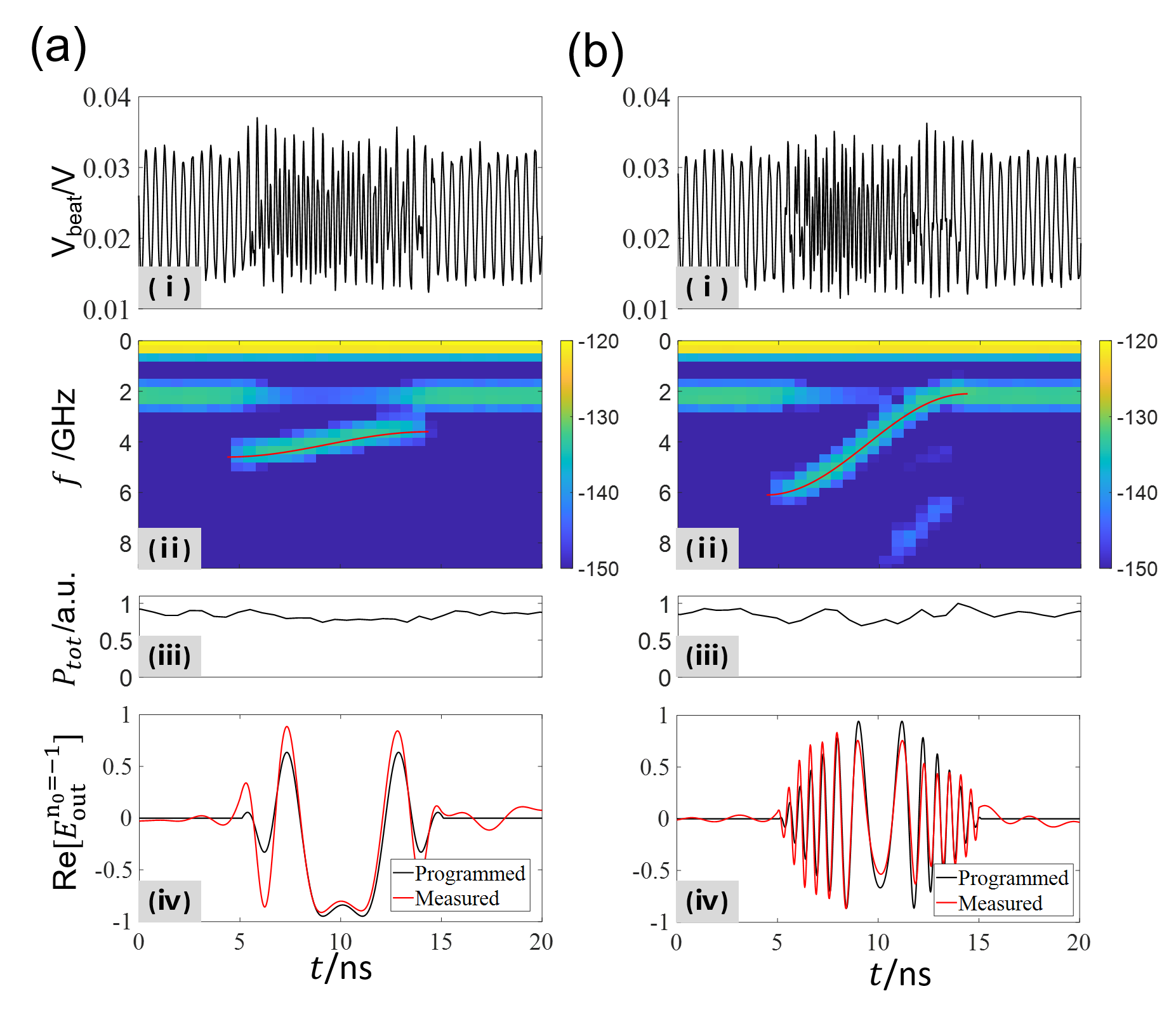}
  \caption{\label{fig:pulseIC} Characterization of chirped pulses from TSA$_2$ output without Filter$_2$. 
  The frequency sweep range $\Delta f$ is 1~GHz and 4~GHz for the Figs.~(a)(b) data,  respectively. The heterodyning beat notes are given in (i), from which we derive the spectrogram (ii) and in-phase quadrature ${\rm Re}[E_{-1}]$ (iv). Due to the self-balanced amplification, the total output power stays approximately unchanged during the pulse modulation as suggested in (iii). }
\end{figure}
The microwave amplitude $A(t)$ and $\varphi(t)$ are programmed according to Eq.~(\ref{eq:invB}) to transfer specific waveforms to light. We then perform beat note measurements to characterize the output waveforms from TSA$_{1,2}$. 
In particular, we mix $E_{\rm out}$ with a strong local field $E_{\rm r}$ through a beamsplitter to measure the interference $s={\rm Re}\left (E_{\rm r}^*E_{\rm out}e^{-i\omega_{\rm r}t} \right)$, where $\omega_{\rm r}=2\pi\times 1.2~$GHz  is the relative frequency shift between the local field and the unmodulated ECDL output. All the measurements are performed with a fast photodetector (Thorlabs PDA8GS) with a 9.5~GHz detection bandwidth. 

Typical beat note measurements for the TSA$_2$ output are given in Fig.~\ref{fig:pulseIC} with the waveforms programmed as chirp pulse modulation according to Eq.~(\ref{eq:invB}) with $C_{-1}(t)= \sin(\pi t/\tau)e^{i \phi_0 \sin(\pi t/\tau)}$ during $0<t<\tau$. Here $\tau=10$~ns pulses are chirped with $\phi_0=10, 40$~rads in Figs.~\ref{fig:pulseIC}(a, i),(b,ii) respectively. The corresponding range of frequency sweep, $\Delta f=\phi_0/\tau$, is 
1~GHz and 4~GHz, respectively. By performing Fourier transform of the beat note data within a shifting Blackman window with a $T_\mathrm{w}=3$~ns width, the beat notes are demodulated into the $f-t$ spectrographs in Figs.~\ref{fig:pulseIC}(a,ii)(b,ii) in log scale. We further plot the target $\dot \phi(t)-t$ curves in red lines onto the spectrographs to demonstrate the accuracy of the frequency-phase control. It is important to note that $E_{\rm out}$ has a constant total power approximately, as in Figs.~\ref{fig:pulseIC}(a,iii)(b,iii), during the full amplitude and phase modulation. The self-balanced output power is a result of balanced $E_{\rm f1,out}$ by the Filter$_1$ to seed TSA$_{1,2}$.  

\begin{figure}[htbp]
  \centering
  \includegraphics[width=0.5\textwidth]{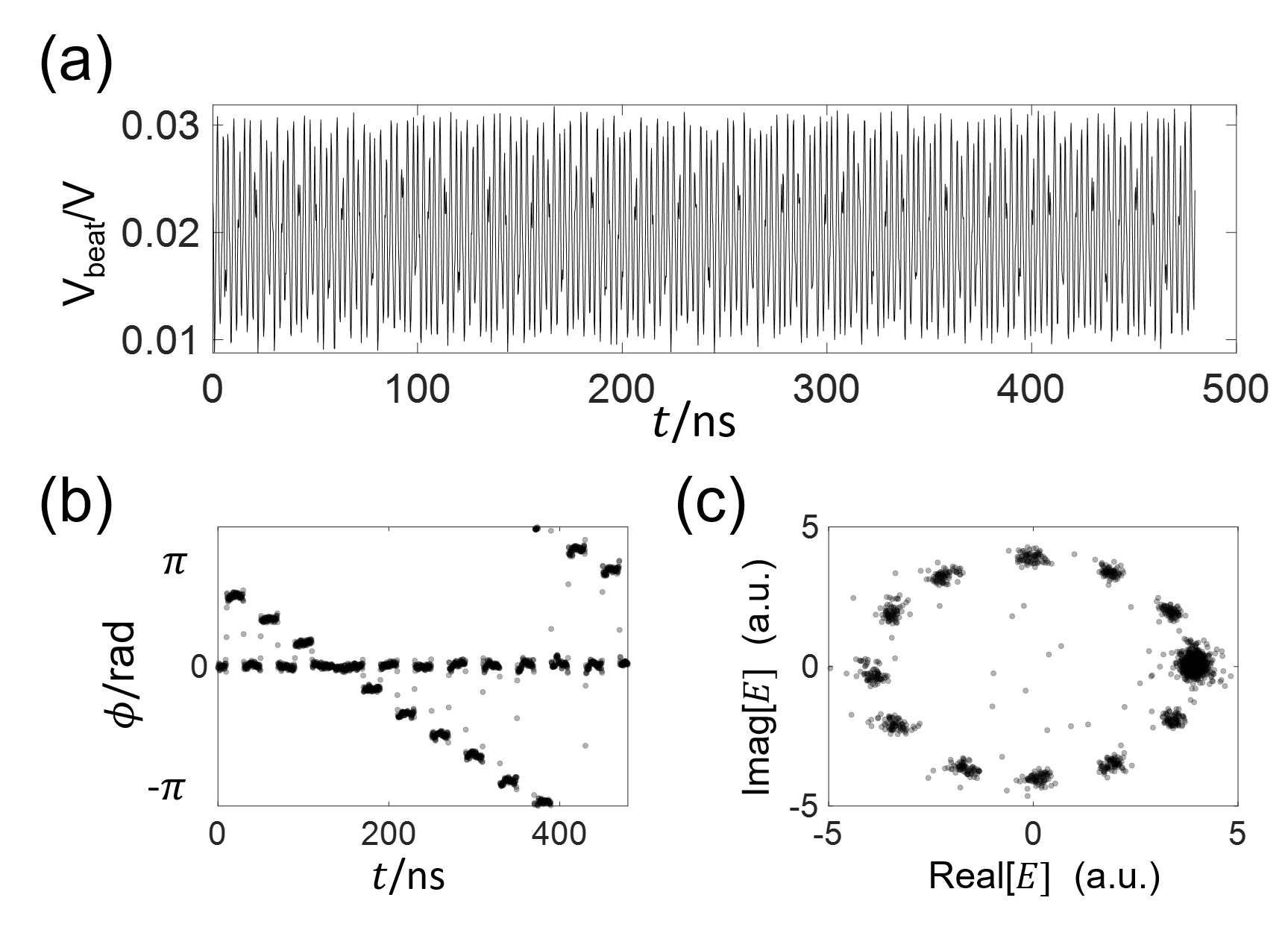}
  \caption{\label{fig:phaseJump} 
Accurate phase modulation of the TSA$_2$ output. The heterodyning beat notes in Fig.~(a) is digitally demodulated as described in the text to obtain the time-dependent phase $\phi(t)$ in Fig.~(b). The complex data is presented in the Fig.~(c) phasor digram.}
\end{figure}

To further confirm the accuracy of the modulated sideband, we use the known target waveform phase $\phi(t)$ to demodulate $E_{-1}(t)$ from Fig.~\ref{fig:pulseIC} (a,i)(b,i) data with a 250~MHz bandwidth. The real parts of $E_{-1}(t)$ are plotted in Figs.~\ref{fig:pulseIC}(a,iv)(b,iv) to compare with the target waveforms. Here we see small deviations of waveform amplitudes from their target values. The deviation is primarily due to the nonlinearity of rf and optical amplification, which can in principle be compensated for by correcting the Eq.~(\ref{eq:invB}) model. On the other hand, the optical frequency and phase are programmed with remarkable accuracy. The accurate phase programmability is further demonstrated in Fig.~\ref{fig:phaseJump} where $\tau=20$~ns pulses are programmed with interleaved phases $\{\phi_{j=2n}=0, \phi_{j=2n+1}=n*\pi/6\}$ for integer $j$ with a constant amplitude. 



\subsection{SPM suppression\label{sec:SPM}}



\begin{figure}[htbp]
  \centering
  \includegraphics[width=0.4 \textwidth]{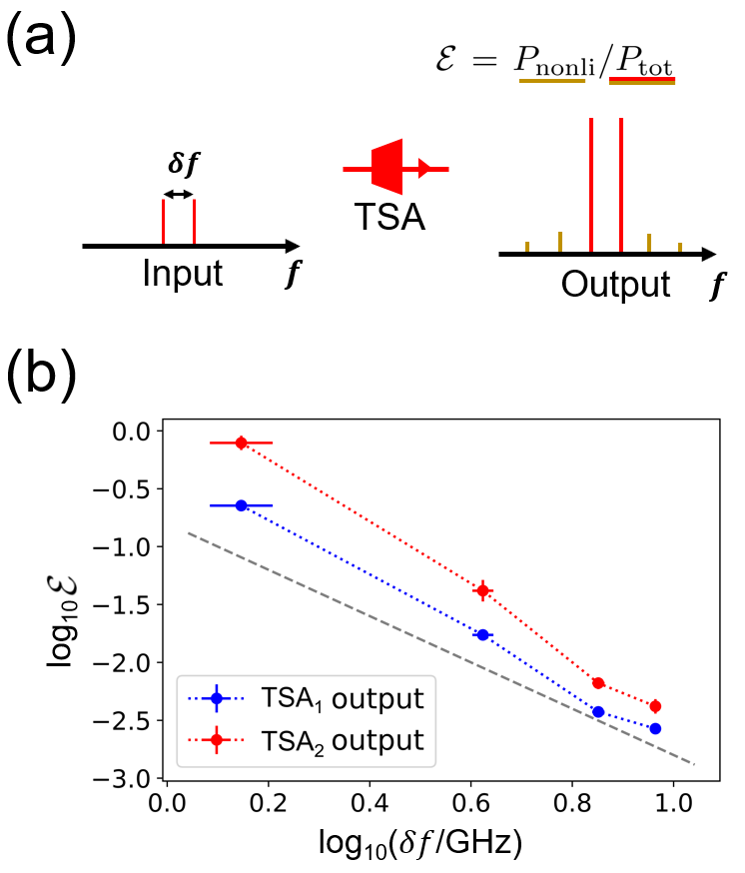}
  \caption{Investigation of SPM as a function of modulation frequency $\delta f$. (a): TSA is seeded with dual-frequency input with frequency difference $\delta f$. The output is analyzed by a Fabry-Perot spectrometer. The nonlinear side band power ratio $\mathcal{E}$ is recorded. (b): Sideband ratio $\mathcal{E}$ vs frequency difference $\delta f$. The error bars reflect the uncertainty of $\delta f$ and $\mathcal{E}$ measurements. The dashed line gives the $1/\delta f^2$ scaling.  
  \label{fig:TA2F}}
\end{figure}

We verify the picture of SPM suppression at large carrier frequency  $\omega_{\rm c}$ with an independent measurement. In particular, we replace the fEOM output in Fig.~\ref{fig:setup-filer} with a seeding laser that contains two frequency components with nearly equal amplitudes and a tunable difference $\delta f$. The total 1~mW power is similar to the fEOM output. As schematically illustrated in Fig.~\ref{fig:TA2F}a,  additional sidebands are generated during the optical amplification through the SPM mechanism. An F-P spectrometer analyzes the power distribution of the sidebands. The SPM efficiency is characterized by the power of the additional sidebands normalized by the total power as $\mathcal{E}=P_{\mathrm{nonli}}/P_{\mathrm{tot}}$. We measure $\mathcal{E}$ as a function of $\delta f$. The results in Fig \ref{fig:TA2F} demonstrate a $1/\delta f^2$-scaling of suppressed SPM at large $\delta f$ , agreeing with similar previous measurements~\cite{Luo2013}. Therefore, for the quasi-continuous seeding by $E_{\rm f1, out}$, the frequency separation of $\delta f=\omega_{\rm c}/2\pi= 6.4~$GHz is large enough that the optical power distributed into additional sidebands are limited to  $\mathcal{E}\approx 2\sim 4\%$ during the optical amplification. 

\subsection{Interleaved cooling and trapping with (in)coherent nanosecond pulses}\label{sec:MixMot}

\begin{figure}[thbp]
  \centering
  \includegraphics[width=0.5\textwidth]{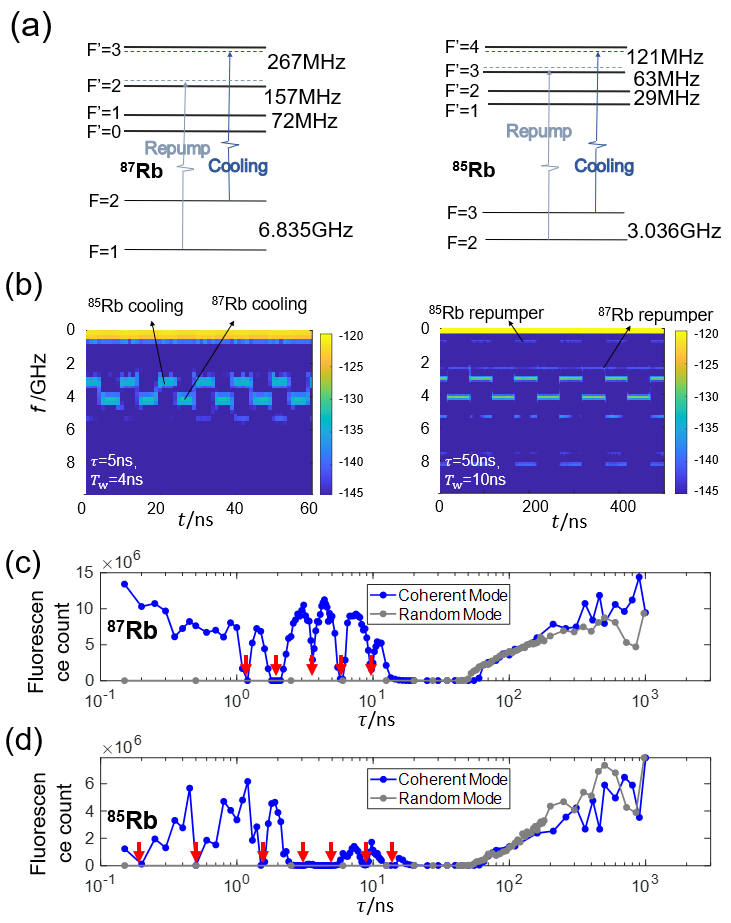}
  \caption{\label{fig:MixMot} (a) The Level diagram and cooling related transitions on the $^{87}$Rb (left) and $^85$Rb (right) D2 line. (b) Spectragraphs derived from heterodyning beat notes of interleaved nanosecond pulses with $\tau$=5~ns (left) and $\tau=50$~ns (right) on log-scale. Fluorescence counts vs $\tau$ for $^{85}$Rb and $^{87}$Rb are shown in Fig.~(c) and Fig.~(d), respectively. Red arrows mark the expected location of $\tau$, where multiple square pulses with $T_{\rm rep}=2\tau$ period and coherent phases resonantly drive hyperfine depumping transitions to degrade the MOT performance. }
\end{figure}

Beyond $^{87}$Rb cooling, the high-power optical waveform generation system is equipped in our lab more generally for cooling, trapping, coherent control and laser spectroscopy of rubidium isotopes (Fig.~\ref{fig:MixMot})~\cite{He2021a,Wang2021}. To address a specific atomic hyperfine transition, we set the microwave modulation frequency $\omega_c$ to the corresponding value and program the slowly varying amplitude and phase of optical pulses according to Eq~(\ref{eq:invB}).  All the D2 transitions of both Rubidium isotopes, as shown in Fig.~\ref{fig:MixMot}a, are able to be generated by the waveform modulation, except for the repumping transition of $^{87}$Rb which is provided by the unmodulated sideband as mentioned before. To simultaneously address multiple transitions , we simply program each component independently according to Eq.~(\ref{eq:invB}). The final voltage signal $V(t)$ to drive the fEOM sums over all the components weighted by the required amplitude ratio, followed by proper normalization to ensure the maximum voltage still optimally drives the -1$^{\rm th}$ sideband in this work.  


Here, we demonstrate the wideband performance of the system by magneto-optical trapping (MOT) with interleaved nanosecond pulses. In particular, microwave pulses with duration $\tau$, carrier frequency $\omega_{c,j}$, amplitude $A_j$, and phase $\phi_j$ are applied with full duty cycle to fEOM in an interleaved fashion to alternatively address $^{85}$Rb and $^{87}$Rb. To address $^{87}$Rb, the carrier frequency $\omega_{c,j}=2\pi\times 6.36$~GHz is chosen as before at even pulse number $j=2n$. To address $^{85}$Rb, $\omega_{c,j}=2\pi\times \{2.32,5.23\}$~GHz are chosen at odd pulse number $j=2n+1$ with two sidebands, with $A_j=\{1,5\}$ for repumping and cooling respectively. A common ``MOT detuning'' from the cooling sidebands to the hyperfine transitions is set as $\Delta_e=-2\pi\times 11$~MHz for both isotopes. The microwave signal $V(t)$ to drive the fEOM is then synthesized as described above. As such, both isotopes are subject to a repetitive train of square pulses with $50\%$ duty cycle at a repetition rate of $f_{\rm rep}=1/2\tau$. Importantly, for $\tau \Gamma_e<1$, the coherent dynamics should be driven by the pulse train. Here the 
lifetime $\tau_e=1/\Gamma_e\approx 26$~ns sets the ``coherent memory time'' for the pulse-to-pulse excitation. To demonstrate the associated dynamics, we sample the pulse duration $\tau$ from 0.15~ns up to 1~$\mu$s, and program the interleaved pulses in a ``coherent mode'' with constant $\phi_j$. For comparison, $\phi_j$ are randomized in a  ``random mode'' to suppress the inter-pulse-driven coherent dynamics.

The amplified TSA$_2$ output waveforms are analyzed with the heterodyning method described in Fig.~\ref{fig:pulseIC}. Examples for $\tau=5$~ns and $\tau=50$~ns are given in Fig.~\ref{fig:MixMot}(b). The unshown heterodyne beat notes are recorded in the ``coherent mode'', although the phase coherence does not appear in the time-resolved spectrum.  The gain saturation by TSAs has not been compensated for during the waveform programming. The SPM suppression is not perfect either. The nonlinearity during the TSA optical amplification leads to weak but visible additional sidebands on the log scale, particularly for the $j=2n+1$ pulses when two seeding sidebands are injected to the TSAs. These additional sidebands hardly impact the operation of the mixed MOT.

The amplified nanosecond pulses with a total power of 700~mW are sent to a double-MOT system where a 2D-plus source MOT feeds a 2nd MOT in the standard 3D configuration. After loading the 2nd MOT for 1~second, we successively take two fluorescence images for $^{85}$Rb and $^{87}$Rb , each with 5~ms exposure time with a CCD camera, by setting the MOT beams resonant to the respective cooling transitions. Typical fluorescence counts are plotted in Figs.\ref{fig:MixMot}(c)(d) as a function of pulse duration $\tau$ for both isotopes. Beyond $\tau=50~$ns, the MOT driven by interleaved pulses behaves similarly in the ``coherent'' and ``random'' modes in terms of atom number, as suggested by fluorescence imaging. The critical role of phase coherence emerges for $\tau<20$~ns where cooling and trapping occur only for coherent pulses.  Here, for the MOT operation, the laser excitations are weak enough in the linear regime, the atomic dynamics is largely decided by the spectrum of the pulses, which forms a frequency comb with $f_{\rm rep}$ periodicity. In particular, as in Fig.~\ref{fig:MixMot}(c), the locations of fluorescence dips for both isotopes are coincide mostly with the spectral analysis, which predicts efficient excitation of ``open'', ``depumping'' transitions by the frequency combs in both isotopes. Without further exploring the cooling scheme in this work, we note that cooling and trapping by the nanosecond coherent pulse train can be an interesting topic for future broadband cooling and trapping~\cite{Wallis1989, Dunning2015, Weitz2000}.



\section{Summary and outlook}
Novel research scenarios in atomic physics and quantum optics~\cite{Miao2007,Jayich2014,Heinrich2018, Long2019, He2020a, Koch2012, Carini2015,Scully2015,Facchinetti2016} require generation of powerful, arbitrarily shapeable optical waveforms with the GHz-level modulation bandwith. A useful strategy has been to optically amplify weak signals from high-speed integrated modulators~\cite{Gould2015, Kaufman2017,Clarke2021}. However, a common problem in this approach is associated with amplification noise and signal distortions, particularly if the signal level is not a constant so that the level of the amplifier's gain-saturation is modulated in time~\cite{AgrawalGovindP1989,Hong1994, Cruz2006, Baveja2010,Luo2013,Meng2018}.


In this work, we have explored a self-balancing technique in amplifying sideband modulation to suppress signal distortion. Sub-milliWatt signal from a fiber-EOM is amplified into Watt-level output with substantially suppressed ASE and SPM noise. The technique exploits relatively slow carrier-relaxation dynamics in TSAs, as has been observed previously~\cite{Luo2013}. We note similar strategy is followed in refs.~\cite{Heinrich2018, Clarke2021} in a pre-designed manner instead of being automatic. For the laser system demonstrated in this work, we remark that the accuracy of the self-balanced seeding as well as the level of TSA saturation drifts on a time-scale of hours to {\it slightly} affect the repeatability of the waveforms. The discrepancies are about a fraction of the distortion observed in Fig.~\ref{fig:pulseIC}. The drifts are likely due to the change of optical injection and the TSA chip working condition.  Improved output waveform stability may be achievable in future work by combining inline optical monitoring with active waveform-corrections.

\section*{Acknowledgement}
We acknowledge support from National Key Research Program of China under Grant No.~2016YFA0302000 and  No.~2017YFA0304204, from NSFC under Grant No.~12074083, and the National Key Scientific Instrument and Equipment Development Project (12027806).

\bibliography{wbl}

 \end{document}